\documentclass[prd,showpacs,amsmath,amssymb,11pt]{revtex4}
\usepackage{latexsym}
\usepackage{bm}

\begin{document}

\title{Comment on the new AdS universe}

\author{{\" O}zg{\" u}r Sar{\i}o\u{g}lu}  
\email{sarioglu@metu.edu.tr}
\affiliation{Department of Physics, Faculty of Arts and  Sciences,\\
             Middle East Technical University, 06531, Ankara, Turkey}

\author{Bayram Tekin}  
\email{btekin@metu.edu.tr}
\affiliation{Department of Physics, Faculty of Arts and  Sciences,\\
             Middle East Technical University, 06531, Ankara, Turkey}

\date{\today}

\begin{abstract}
We show that Bonnor's new Anti-de Sitter (AdS) universe and its $D$-dimensional
generalization is the previously studied AdS soliton.
\end{abstract}

\pacs{04.20.-q, 04.20.Jb, 04.50-h}

\maketitle

Recently, Bonnor \cite{bonnor} noted a second non-singular solution
to cosmological Einstein equations $R_{\mu\nu} = \Lambda g_{\mu\nu}$ in 
\textit{four} dimensions, with $\Lambda < 0$; the first obviously being the 
maximally symmetric AdS spacetime. In this comment, we show that Bonnor's 
``non-uniform AdS universe'' is nothing but the AdS soliton, whose higher 
dimensional generalizations also exist \cite{horo}.

The metric introduced in \cite{bonnor} belongs to a family of cylindrically symmetric,
static spacetimes that are solutions to the four dimensional Einstein equations 
with a negative cosmological constant, and that were independently found a long time 
ago by Linet \cite{linet} and Tian \cite{tian}. Various properties of these
metrics were studied in \cite{zofbic}, where non-singular sheet sources of them 
were also found and their relation to the black-string solutions were examined.
Roughly speaking, these can be given the interpretation of infinite rods whose
mass per unit length is defined through a parameter $\sigma$ with a specific range.
The spacetime that Bonnor considered is the one obtained by setting
this parameter $\sigma = 0$ (see \cite{zofbic} and \cite{bonnor} for details). 

The generalization of this solution to $D=n+1$ dimensions follows as
\begin{equation}
ds^{2} = d \rho^{2} + \cosh^{4/n}\Big( \frac{n \rho}{2\ell} \Big) \, \Big(-dt^{2} + 
\frac{4 \ell^{2}}{n^{2}} \, \tanh^{2}\Big( \frac{n \rho}{2\ell} \Big) \, d \phi^{2} +
\sum_{i=1}^{n-2} dx_{i}^{2}\Big) \,,
\label{newads}
\end{equation}
which solves
\[ R_{\mu\nu} = - \frac{n}{\ell^{2}} g_{\mu\nu} \, , \]
with the ranges \( t, x_{i} \in \mathbb{R} \) (where $i=1,\dots, n-2$), 
\( \rho \in \mathbb{R}^{+} \) and \( \phi \in [0,2\pi) \). When $n=3$, this 
reduces to the metric in \cite{bonnor}. For $n=2$, one simply gets the AdS 
solution in $2+1$ dimensions. (\ref{newads}) has no singularities or horizons. 
Any invariant that involves only the curvature scalar or the Ricci tensor is 
identical to its counterpart belonging to the usual maximally symmetric AdS. 
However, the regularity and the non-uniformity of the solution is apparent from 
the Kretschmann scalar
\begin{equation}
R_{\mu\nu\lambda\sigma} R^{\mu\nu\lambda\sigma} = \frac{1}{\ell^{4} \cosh^{4} \Big( \frac{n \rho}{2\ell} \Big)}
\Big( n(n-2)(n-1)^{2} + \frac{n(n+1)}{2} \Big[ 1+ \cosh\Big( \frac{n \rho}{\ell} \Big) \Big]^{2} \Big) \, .
\end{equation}

Let us show that (\ref{newads}) is a special case of the AdS soliton \cite{horo}, which reads
\begin{equation}
ds^{2} = \frac{r^2}{\ell^2} \left[ \left( 1 - \frac{r_0^{n}}{r^{n}}
\right) d\tau^2 + \sum_{i=1}^{n-2} (dx^i)^2 - dt^2 \right] + 
\left( 1 - \frac{r_0^{n}}{r^{n}} \right)^{-1} \, \frac{\ell^2}{r^2} \, dr^2 \,.
\label{adssoliton}
\end{equation}
(\ref{adssoliton}) was obtained by the double analytic continuation of a near extremal 
$(n-1)$-brane solution. $x^i$ and the $t$ denote the coordinates on the ``brane"  and 
$r \ge r_0$, where $r_{0}$ is a free parameter. A conical singularity is avoided if
$\tau$ has a period \( \beta = 4 \pi \ell^2 / (n r_0) \) \cite{horo}. 

Consider the coordinate transformation \( r = r_{0} \cosh^{2/n}{(n \rho/2 \ell)} \),
which respects the proper ranges of $r$ and $\rho$ \footnote{This transformation
was suggested by Roberto Emparan.}. Then (\ref{adssoliton}) turns into
\begin{equation}
ds^{2} = d \rho^{2} + \frac{r_{0}^{2}}{\ell^{2}} \, \cosh^{4/n}\Big( \frac{n \rho}{2\ell} \Big) \, 
\Big(-dt^{2} + \tanh^{2}\Big( \frac{n \rho}{2\ell} \Big) \, d \tau^{2} +
\sum_{i=1}^{n-2} dx_{i}^{2}\Big) \,.
\label{adssol2}
\end{equation}
Choosing $r_{0}=\ell$ and $\tau=2 \ell \phi/n$, one obtains (\ref{newads}).

Finally, it is obvious that (\ref{newads}) does not make sense in $D=1+1$ 
dimensions. However, one can separately consider the $D=2$ Euclidean space. 
There, the cigar soliton of Witten \cite{witten} 
\begin{equation}
ds^{2} = d \rho^{2} + \tanh^{2}{\rho} \, d \phi^2
\label{cigar}
\end{equation}
can be thought of as the $D=2$ analog of (\ref{newads}). 
Note that, the cigar is also singularity free with a curvature scalar
\( R = 4/\cosh^{2}{\rho} \) but, of course, it is asymptotically flat
unlike (\ref{newads}).

To conclude, we have shown that Bonnor's solution \cite{bonnor} and its $D$-dimensional
generalization is the AdS soliton \cite{horo}.

\section{\label{ackno} Acknowledgments}
In the initial version of this note, we weren't aware that Bonnor's solution
is a special case of the AdS soliton. We kindly thank Roberto Emparan who
brought this to our attention. That led to a complete revision of our work. 
This work is partially supported by the Scientific and Technological Research
Council of Turkey (T{\"U}B\.{I}TAK). B.T. is also partially supported by
the T{\"U}B\.{I}TAK Kariyer Grant 104T177.

\end{document}